# Thermal Conductance of an Individual Single-Wall Carbon Nanotube above Room Temperature

Eric Pop,[1,2] David Mann,[1] Qian Wang,[1] Kenneth Goodson[2] and Hongjie Dai[1,*]

[1] *Department of Chemistry and Laboratory for Advanced Materials, Stanford University, Stanford, CA 94305, USA*

[2] *Department of Mechanical Engineering and Thermal Sciences, Stanford University, Stanford, CA 94305, USA*

The thermal properties of a suspended metallic single-wall carbon nanotube (SWNT) are extracted from its high-bias (*I-V*) electrical characteristics over the 300–800 K temperature range, achieved by Joule self-heating. The thermal conductance is approximately 2.4 nW/K and the thermal conductivity is nearly 3500 Wm$^{-1}$K$^{-1}$ at room temperature for a SWNT of length 2.6 μm and diameter 1.7 nm. A subtle decrease in thermal conductivity steeper than 1/*T* is observed at the upper end of the temperature range, which is attributed to second-order three-phonon scattering between two acoustic modes and one optical mode. We discuss sources of uncertainty and propose a simple analytical model for the SWNT thermal conductivity including length and temperature dependence.

[*] E-mail: hdai@stanford.edu

Single-wall carbon nanotubes (SWNTs) have sparked great scientific and engineering interest owing to their outstanding electrical and thermal properties. Consequently, they have been proposed for applications in integrated circuits (as transistors or interconnects)[1-3] and in thermal management (as thermal interface materials).[4] In both cases, knowledge of their thermal properties is key to understanding their overall behavior.[5] While several theoretical studies exist on the thermal conductivity of individual SWNTs,[6-11] only few experimental estimates are currently available,[12] and no investigations exist above room temperature. However, it is this temperature range that is expected to be of most practical importance for applications of carbon nanotubes in electronics and integrated circuits.

In this Letter we extract the thermal conductance of an individually suspended SWNT in the 300–800 K temperature range. For the first time, we empirically find a subtle decrease in the thermal conductivity of the SWNT near the upper end of this temperature range, which is proportional to $1/T^2$ due to second-order three-phonon scattering processes. The approach presented employs direct (DC) self-heating of the nanotube under high-bias current flow in several ambient temperatures (from 250–400 K). This method relies on the direct relationship between the SWNT lattice temperature and phonon-limited electrical transport, and may be generally applicable to other one-dimensional systems where similar transport conditions occur at high electrical bias (e.g., nanowires or other molecular wires).

Suspended SWNTs were grown across Pt contacts and lithographically pre-defined trenches, as described elsewhere.[13,14] The nanotubes were characterized with SEM and AFM to ensure single-tube connections and to obtain length ($L$) and diameter ($d$) information. The SEM image of a typical suspended SWNT is shown in Fig. 1. Electrical characterization was done in vacuum (approximately $10^{-6}$ torr), at ambient temperatures between 250 and 400 K. The measured *I-V* characteristics of a metallic suspended SWNT with $L \approx$ 2.6 μm and $d \approx$ 1.7 nm are displayed in Fig. 2 (symbols). The thermal isolation of the tube (no substrate or ambient gas) ensures that all heat produced from Joule self-heating is conducted along the tube itself (heat loss by blackbody radiation is estimated to be at most a few nano-Watts, whereas the dissipated Joule power is three orders of magnitude greater, at several micro-Watts). This is essential for the high-bias SWNT behavior and for extracting its thermal conductance, as explained below. Unlike SWNTs on substrates which carry more than 20 μA current,[15, 16] suspended SWNTs exhibit Negative Differential Conductance (NDC) and a current-carrying ability almost an order of magnitude lower at high bias, when significant self-heating and non-equilibrium phonons strongly impede electrical transport.[5]

Electrical transport in the low-bias region of the *I-V* curves in Fig. 2 is limited by the finite resistance of the contacts ($R_C \approx$ 10 kΩ) and by electron scattering with acoustic (AC) phonons, with a mean free path $\lambda_{AC} \approx$ 1.6 μm at room temperature.[5, 16] Self-heating in this region is not significant, and the observed *I-V* characteristics are similar to those of SWNTs on substrates. The electrical resistance of the suspended SWNT under high-bias ($V > 0.3$ V) is however more than an order of magnitude higher (~ 300 kΩ at 1 V in Fig. 2), and limited by Joule self-heating and strong electron scattering with high energy optical phonons (OPs).[5] This behavior enables our extraction of suspended SWNT thermal properties at high temperature directly from the electrical characteristics at high bias.

We model the resistance of the SWNT with the Landauer-Büttiker approach:[3, 5]

$$R(V,T) = R_C + \frac{h}{4q^2} \frac{L + \lambda_{eff}(V,T)}{\lambda_{eff}(V,T)} \quad (1)$$

where $R_C$ is the electrical contact resistance and $\lambda_{eff} = (1/\lambda_{AC} + 1/\lambda_{OP,ems} + 1/\lambda_{OP,abs})^{-1}$ is the effective electron mean free path (MFP) which includes elastic scattering with AC phonons, and inelastic OP emission and absorption. The MFPs are dependent on temperature and bias, as explained in detail elsewhere.[3,5] In order to obtain the *I-V* curves, the scattering MFPs and consequently the SWNT resistance are computed self-consistently with the temperature and the Joule power dissipated along the nanotube, using the heat conduction equation:

$$A\nabla(k\nabla T) + I^2(R - R_C)/L = 0 \qquad (2)$$

where the second term is the Joule heating per unit length and $A=\pi d b$ is the cross-sectional area ($b$ = 3.4 Å the tube wall thickness). Naturally, a key component of this approach is the thermal conductivity $k$ along the SWNT. We have found that we can pose the inverse problem, asking what thermal conductivity is necessary (for a given bias, power dissipation, and hence a given SWNT temperature profile) to reproduce the experimentally measured *I-V* curves. Since the suspended tube suffers significant self-heating only at high bias, we focus specifically on the voltage region $V > 0.3$ V.[17] To gain a qualitative understanding, we note that transport is strongly limited by OP scattering in this regime, and the resistance becomes simply inversely proportional to the net OP scattering MFP alone: $R \sim (h/4q^2)(L/\lambda_{OP})$. This MFP scales inversely with the OP occupation ($N_{OP}$), which is proportional to temperature in this range.[5] We can summarize these relations as $R \propto 1/\lambda_{OP} \propto N_{OP} \propto T$. Here, the temperature is determined by the Joule power dissipated along the nanotube and by its thermal conductivity $k$, as in Eq. 2. This provides physical insight into the electro-thermal behavior of the SWNT at high bias and explains how the thermal conductivity is directly linked to the measured high bias electrical characteristics.[5] Using the high bias region of the *I-V* curves is also essential because it minimizes the role of other parameter variations (e.g., due to imperfect knowledge of $R_C$ or $\lambda_{AC}$) and hence of errors associated with them.

A schematic block diagram of our inverse thermal conductivity extraction procedure is shown in Fig. 3. Quantitatively, we compute the "forward solution" $R(V,T)$ as in Eq. 1 and Ref. 5, and consider individual bias points along the *I-V* curves in Fig. 2. Given the experimentally measured resistance and power at a given applied bias, we adaptively modify the thermal conductivity input (in Eq. 2) of our self-consistent model until the agreement with the measured *I-V* data is within 0.5 percent. This is done by starting with an initial guess $k_0 = 3000$ Wm$^{-1}$K$^{-1}$, then calculating the theoretically expected current for the starting bias ($V_0 = 0.3$ V). If this current is larger (smaller) than the experimental value, then the value of $k$ is decreased (increased), and the temperature and current are reevaluated. The procedure is repeated until good agreement between the calculated and measured current is obtained. The converged solution is then used as the initial guess input for the next bias point. The series of computed current values creates the solid lines plotted in Fig. 2, the agreement with the data being essentially "perfect" by design. Given the power input and the *average* SWNT temperature estimate at every bias point, a plot of this best-fit thermal conductivity vs. average temperature can be obtained, as shown in Fig. 4. Having carried out the electrical characterization at various ambient temperatures from 250–400 K, the thermal conductivity dependence on temperature under Joule heating is therefore deduced approximately from 300 to 800 K. We note that the extracted trend in thermal conductivity is very clear, despite (or perhaps owing to) the straightforward nature of this approach. That the four extracted data sets in Fig. 4 line up consistently well is also very encouraging, given that the only input parameter changed between extractions was the ambient temperature $T_0$.

At first glance, the observed trend of the high temperature thermal conductivity in Fig. 4 is consistent with Umklapp phonon-phonon scattering[18] which gives an approximately $1/T$ temperature dependence, as was suggested previously.[5] However, a more subtle effect can be noted at the upper end of the spanned temperature range, i.e. a drop of the thermal conductivity at a rate steeper than $1/T$. This is attributed to second-order three-phonon scattering processes,[10, 19] whose scattering rates are proportional to $T^2$, leading to a thermal conductivity that scales as $1/(\alpha T + \beta T^2)$ where $\alpha$ and $\beta$ are constants.[20] Such behavior has been previously observed, e.g., in the experimental Si and Ge thermal conductivity at high temperatures,[21] but not in SWNTs until now. In the case of our SWNT, the three-phonon scattering process very likely involves the anharmonic interaction of two acoustic modes and one optical mode,[20] since the population of the latter is known to be significantly raised in suspended metallic SWNTs at high temperature and bias.[5] At the low end of the temperature range spanned by this study (near room temperature) we observe a leveling of the thermal conductivity, suggesting a transition towards thermal transport limited by phonon-boundary scattering due to our finite sample size. We ought to note that in the examined temperature range the thermal conductivity of the SWNT is owed to transport via phonons, the electronic contribution being negligibly small.[8]

The extracted value of the thermal conductance for the SWNT examined here is approximately $G \approx 2.4$ nW/K, and the thermal conductivity is $k \approx 3500$ Wm$^{-1}$K$^{-1}$ at room temperature for the length $L \approx 2.6$ μm and diameter $d \approx 1.7$ nm. The thermal conductivity is found to decrease to approximately 1000 Wm$^{-1}$K$^{-1}$ at 800 K. We note that the thermal characteristics extracted here must rely on an assessment of the SWNT temperature, which is very difficult to obtain experimentally in a quantitative manner.[22] However, the tempera-ture under which SWNTs break down when exposed to oxygen is relatively well established. This is the correlation we have exploited[5] to obtain a calibration of our approach, and specifically of Eq. 2. It is known that when exposed to air (oxygen) SWNTs begin to break by oxidation at 800–900 K.[23] We have used suspended SWNTs of similar length and diameter, and found breakdown voltages consistently between 2.3–2.5 V in 10 torr oxygen ambient. The relatively low pressure of the oxygen ambient is necessary to ensure all heat transport mechanisms are similar to those in vacuum (i.e. negligible heat loss to gas), yet the presence of the oxygen guarantees breakdown by oxidation, this reaction temperature being relatively well known as mentioned above. We can now perform an estimate of the error introduced in the thermal conductivity extraction given the uncertainty in the burning temperature. Assuming a maximal, worst-case 800–1000 K range for the burning temperature, we find a 2800–3900 Wm$^{-1}$K$^{-1}$ range in the extracted thermal conductivity around room temperature, and a narrower 1000–1160 Wm$^{-1}$K$^{-1}$ range in the extracted conductivity at 800 K in vacuum. However, even considering both extremes, the decrease in thermal conductivity with temperature is consistently found to be steeper than $1/T$, suggesting that even the widest possible range of burning temperatures cannot rule out our observation of second-order three-phonon scattering at very high temperatures. On the other hand, we note that aside from this paragraph's discussion, in the rest of the paper (in the figures, etc.) a burning temperature of 600 °C = 873 K was used for the numerical extraction of thermal parameters.

In order to compare our results with others currently available in the literature[12] we plot the diameter-adjusted thermal conductivity in Fig. 5 as $kd = GL/\pi b$, where $G$ is the thermal conductance. In this work, we define a SWNT cross-sectional area which scales linearly with diameter (discussion surrounding Eq. 2 above), consistently with Refs. 10 and

12. Near room temperature our results are within 20 percent of the other data set currently available.[12] The discrepancy can be attributed (in increasingly likely order of importance) to differences in length[10] ($L \approx 2.8$ μm in Ref. 12 and 2.6 μm in this work), chirality, or thermal contact resistance. Differences in chirality alone could lead up to a 20 percent variation in thermal conductivity, based on some theoretical estimates.[11] Regarding the thermal resistance of the contacts, we have estimated it as follows. The length of our nanotube-Pt contacts is approximately $L_C \approx 2$ μm, and given the nanotube diameter we estimate the contact area $A_C = dL_C$. Typical values of the thermal resistance per unit area for a wide array of measured solid material interfaces all fall in the range of $1-3 \times 10^{-8}$ m$^2$K/W at room temperature.[24] Hence, we expect the thermal contact resistance between our SWNT and the Pt pad to fall in the range of $3-9 \times 10^6$ K/W. We have assumed a thermal resistance at the contacts of $\mathcal{R}_{C,th} = 6 \times 10^6$ K/W in our extraction results displayed in Figs. 3 and 4. The imprecision in our knowledge of this value can lead to an uncertainty of approximately 10 percent in the extracted SWNT thermal conductivity. We find that a somewhat larger value $\mathcal{R}_{C,th} \approx 1.6 \times 10^7$ K/W would be needed to extract a room temperature thermal conductance comparable to that of Ref. 12 (assuming that chirality and length differences are negligible). This is not outside the realm of possibility, given that the highest measured interface thermal resistance (upper end of the range mentioned above) was across a Pb/diamond interface,[25] which may be of comparable thermal quality with our Pt/SWNT interface. Although exact conclusions are difficult to draw, this discussion nevertheless highlights that knowledge of the thermal contact resistance is important for determining the intrinsic thermal properties of a nanostructure.

The length dependence of the thermal conductivity can be understood since the phonon MFP at room temperature is approximately 0.5 µm.[12] If the sample size is comparable to or shorter than this length, the thermal conductivity is limited by phonon-boundary scattering, rather than by phonon-phonon scattering processes. This leads to an apparent lower and less temperature-sensitive thermal conductivity for shorter nanotubes than longer ones.[10] We propose a simple, analytical model for the length and temperature dependence of SWNT thermal conductivity which can be used for quick future estimates:

$$k(L,T) = \left[3.7 \times 10^{-7} T + 9.7 \times 10^{-10} T^2 + 9.3(1 + 0.5/L)T^{-2}\right]^{-1} \quad (3)$$

where $k$ is in units of $Wm^{-1}K^{-1}$, the length $L$ is in microns and $T$ is in Kelvins. This is obtained as a fit to the combined low- and high-temperature data presented in Fig. 5. Similar expressions have been previously used to reproduce the thermal conductivity of metals over a wide temperature range.[26] The length enters the formula through a simple Matthiessen's Rule estimate,[18] where 0.5 is the room-temperature phonon MFP in microns, and the other constants are fitting parameters. The simple formula properly reproduces the thermal conductivity behavior over the wide 100–800 K temperature range, but must be modified at lower temperatures. Note that at high temperature the $T^{-2}$ term vanishes and the expression follows the expected $1/(\alpha T + \beta T^2)$ trend, as previously discussed. The dependence on sample size ($L$) is also eliminated at high temperatures, when the thermal conductivity is limited by phonon-phonon scattering rather than by scattering with the sample boundaries. Using Eq. 3, we plot the estimated thermal conductivity vs. temperature for SWNTs of various lengths in Fig. 6, demonstrating sensible agreement with results of more sophisticated simulations[10] and with the experimentally observed behavior discussed earlier. Future work must evidently evaluate samples with different lengths to obtain a more comprehen-sive understanding of SWNT thermal conductivity. Nevertheless, the present study along with that of Ref. 12 provide the first empirical picture of the SWNT thermal conductivity in the 2.6–2.8 μm length range and the 100–800 K temperature range.

In summary, we have presented a method for extracting the thermal conductivity of an individual SWNT from high-bias electrical measurements by careful reverse fitting using an existing electro-thermal transport model. The values are obtained over the 300–800 K range, suitably complementing the only other data set currently available[12] and representing the first high temperature study of SWNT thermal properties. The thermal conductivity is found to decrease more steeply than the expected $1/T$ at the upper end of the temperature range, a behavior attributed to second-order three-phonon processes with a scattering time proportional to $1/T^2$. The method presented here may be generally applicable to extracting thermal properties of other one-dimensional structures where similar transport conditions occur under self-heating at high bias.

We acknowledge valuable technical discussions with Natalio Mingo and Vincent Terrapon, feedback on an earlier manuscript from Matthew Panzer, and financial support from the Marco MSD Center.

**Figure Captions:**

**Figure 1**: Scanning electron microscope (SEM) image of a typical SWNT freely suspended across a 2 μm trench and lying on top of the Pt contacts. This sample was coated with 1.5 nm Ti/2.5 nm Au to facilitate SEM imaging.

**Figure 2**: Electrical (current vs. voltage) characteristics of a suspended SWNT with $L \approx 2.6$ μm and $d \approx 1.7$ nm. The symbols represent the experimental data measured in vacuum at ambient temperatures $T_0$ from 250 to 400 K; the lines are the results of our calculations using best-fit values of thermal conductivity at high temperature and high bias (Fig. 4). Not all data points taken are shown, for clarity.

**Figure 3**: Block diagram of the inverse numerical method used for the extraction of thermal conductivity from experimental *I-V* data. The details of the forward solver $I(V,T)$ are given in more detail in Ref. 5. The thermal conductivity at each bias and average SWNT temperature is adjusted until good agreement is obtained with the experimental data. Consequently, a sequence of $(k,T)$ pairs are extracted from bias points above $V_0 = 0.3$ V.

**Figure 4**: Extracted values of the thermal conductivity vs. average SWNT temperature from fitting the high bias *I-V* data in Fig. 2. The symbols used are consistent with those in Fig. 2, squares corresponding to data taken in $T_0 = 400$ K ambient, and so on. The dashed line marks the $1/T$ trend expected due to Umklapp phonon-phonon scattering.[5] We note an additional, more subtle effect at high temperature due to three-phonon interaction between one optical and two acoustic modes, with a scattering time and contribution to thermal conductivity proportional to $1/T^2$.

**Figure 5**: Comparison of diameter-adjusted thermal conductivity ($kd = GL/\pi b$) between our extracted data set (above room temperature) and that of Ref. 12 (below room temperature). The adjustment is necessary to circumvent lack of good diameter knowledge, as both approaches first extract the thermal conductance *G*. In addition, differences and uncertainties

in chirality and thermal contact resistance contribute to the observed discrepancy. An analysis of the magnitude and effects of these sources of uncertainty is provided in the text.

**Figure 6**: Analytic plot of the intrinsic SWNT thermal conductivity over the 100–800 K temperature range as computed with Eq. 3. The length dependence is included heuristically with a simple scaling argument, but differences in chirality may lead to variations up to 20 percent between different tubes.[11]

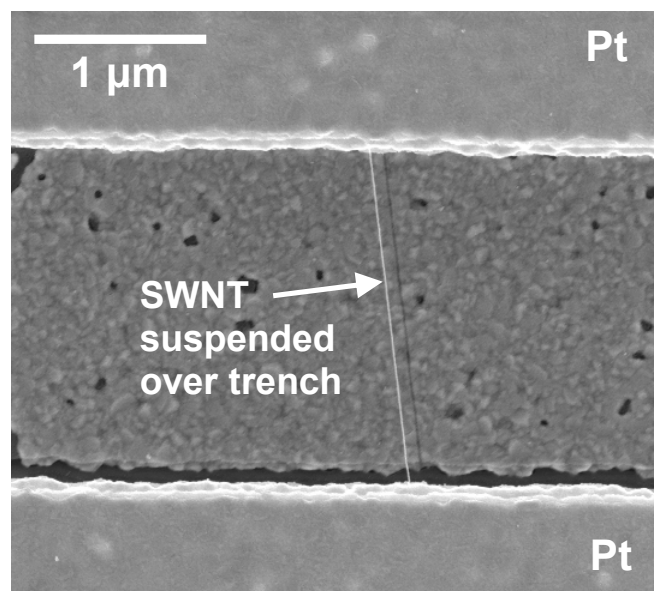

Figure 1.

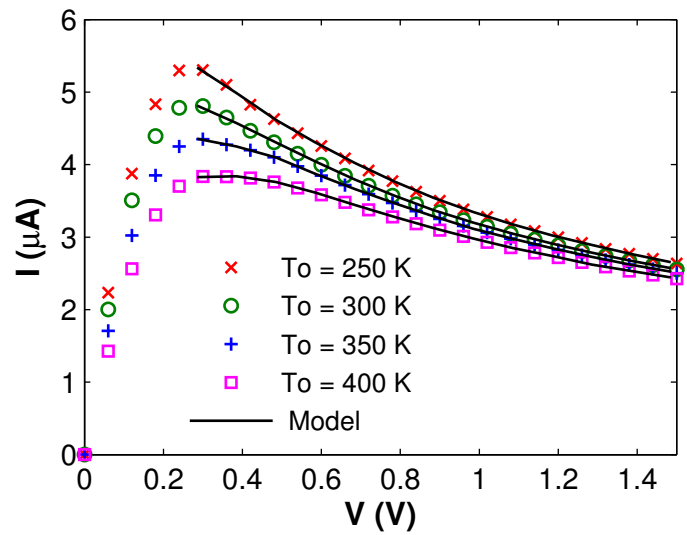

Figure 2.

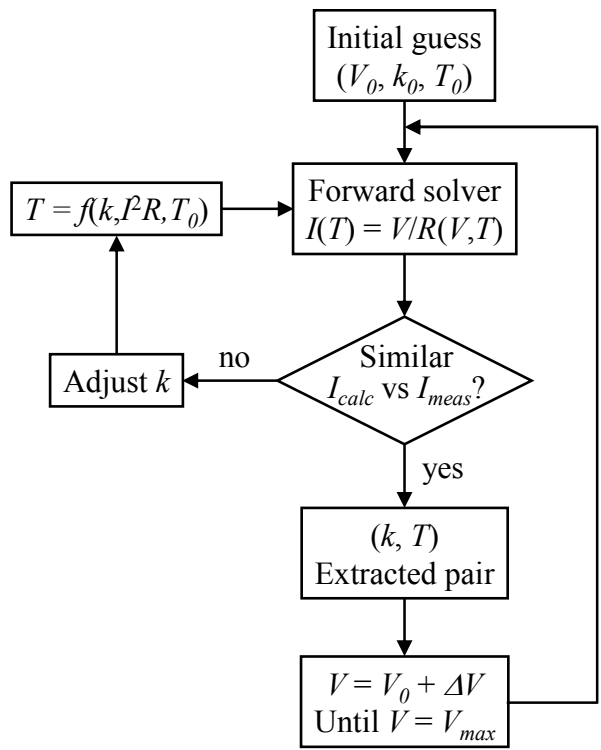

Figure 3.

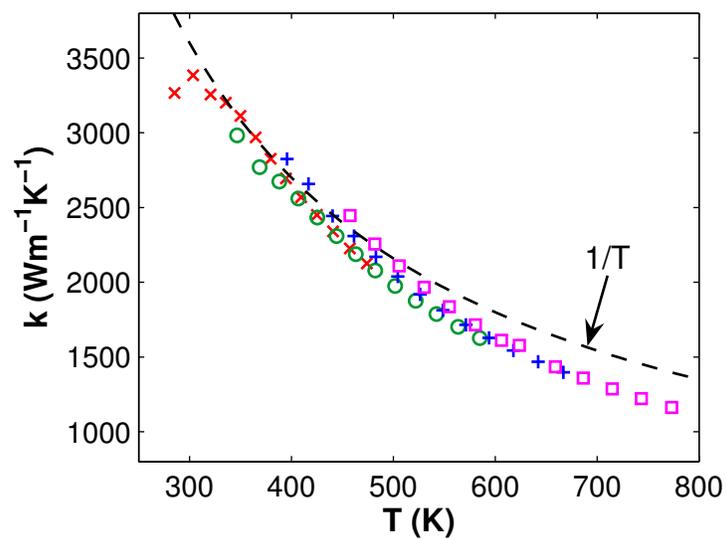

Figure 4.

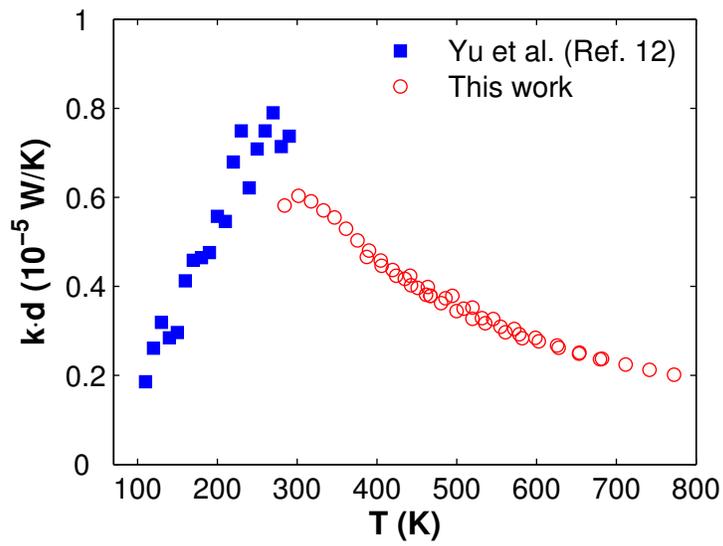

Figure 5.

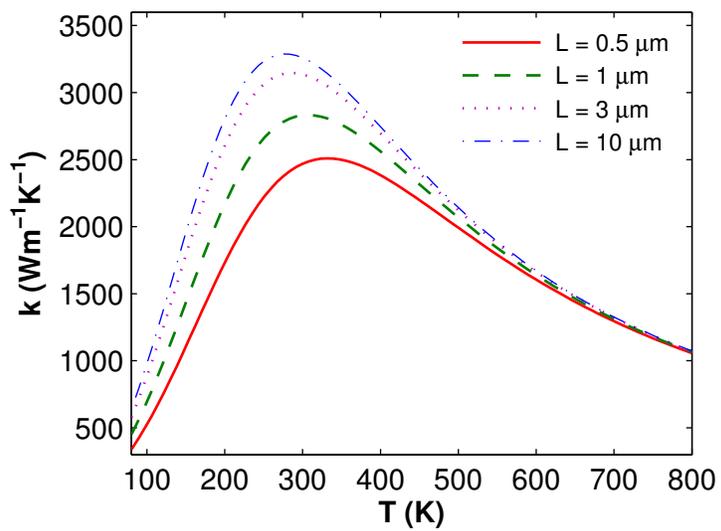

Figure 6.